\documentclass[aps,showpacs,superscriptaddress,preprint]{revtex4}
\usepackage{amsmath}
\usepackage{epsf}
\usepackage{epsfig}

\begin{document}
\title{Joint Entropy of the Harmonic Oscillator with Time Dependent
Mass and Frequency }
\author{Ethem AKT\"{U}RK}
\email[E-mail: ]{eakturk@hacettepe.edu.tr}\affiliation{Department
of Physics, Hacettepe University, 06800, Ankara,Turkey}
\author{\"{O}zg\"{u}r \"{O}ZCAN}
\email[E-mail: ]{ozcano@hacettepe.edu.tr}\affiliation{Department
of Physics Education, Hacettepe University, 06800, Ankara,Turkey}
\author{Ramazan SEVER}
\email[E-mail: ]{sever@metu.edu.tr}\affiliation{Department of
Physics, Middle East Technical University, 06800, Ankara,Turkey}
\begin{abstract}
Time dependent entropy of harmonic oscillator with time dependent
mass and frequency are investigated. The joint entropy so called
Leipnik's entropy is calculated by using time dependent wave
function obtained by the Feynman path integral method. It is shown
that, Leipnik's entropy fluctuates with time. However in constant
mass and time dependent frequency case, entropy increases
monotonically with time.

\end{abstract}

\maketitle

\section{Introduction}
The information entropy plays a major role in a stronger
formulation of the uncertainty relations~\cite{ekrem}. This
relation may be mathematically defined by using the
Boltzmann-Shannon information entropy and the von Neumann entropy.
For both open and closed quantum systems, different
information-theoretical entropy measures have been
discussed~\cite{Zurek,Omnes,Anastopoulos}. The joint
entropy~\cite{Leipnik,Dodonov} can also be used to explain the
properties of the loss of information evolving pure quantum
states~\cite{Trigger}. The joint entropy of the physical systems
were conjectured by Dunkel and Trigger~\cite{Dunkel} in which
their systems named MACS (maximal classical states). The Leipnik
entropy of the simple harmonic oscillator was determined not
monotonically increase with time~\cite{Garbaczewski}. In this
work, we give a uniform description of the complete joint entropy
information  for time dependent entropy of harmonic oscillator
with time dependent mass and frequency. The study of harmonic
oscillators with time dependent frequencies or with time dependent
masses (or both simultaneously) has attracted considerable
interest in past few years but the investigation of joint entropy
of this system have not been studied yet. The time-dependent
harmonic oscillator has invoked much attention because of its many
applications in different areas of physics, such as quantum optics
and plasma physics~\cite{Dandas,Abdalla,Lemos,Ben}.

This paper is organized as follows. In section II, we explain
fundamental definitions  needed for the calculation. In section
III, we get the results for harmonic oscillator with time
dependent mass and frequency. Moreover, we obtain the analytical
solution of kernel and  using this, wave function in both
coordinate and momentum space and its joint entropy were
calculated. In this section, we also investigated same quantities
for harmonic oscillator with strongly pulsating mass and inverse
square time dependent frequency. Finally, we present the
conclusion in section IV.

\section{Fundamental Definitions}
We consider a classical system with $d=sN$ degrees of freedom,
where N is the particle number and s is number of spatial
dimensions~\cite{Dunkel}. The density function
$g(x,p,t)=g(x_1,...,x_d,p_1,...,p_d,t)$ which is non-negative,
time dependent phase space density function of the system is
assuming to be normalized to unity,
\begin{equation}
\int dx dp g(x,p,t)=1.
\end{equation}
The Gibbs-Shannon entropy is described by
\begin{equation}
S(t)=-\frac{1}{N!}\int dx dp g(x,p,t)ln(h^{d} g(x,p,t)),
\end{equation}
where $h=2\pi\hbar$ is the Planck constant. Schr\"{o}dinger wave
equation with the Born interpretation~\cite{Born} is given by
\begin{equation}
i\hbar\frac{\partial\psi}{\partial t}=\hat{H}\psi.
\end{equation}
The quantum probability densities are defined in position and
momentum spaces as $|\psi(x,t)|^2$ and $|\tilde{\psi}(p,t)|^2$,
where $|\tilde{\psi}(p,t)|^2$ is given  as
\begin{equation}
    \tilde{\psi}(p,t)=\int\frac{dx
    e^{-ipx/\hbar}}{(2\pi\hbar)^{d/2}}\psi(x,t).
\end{equation}
Leipnik proposed the product function as~\cite{Dunkel}.
\begin{equation}
    g_{j}(x,p,t)=|\psi(x,t)|^2|\tilde{\psi}(p,t)|^2\geq0.
\end{equation}
Substituting Eq. (5) into Eq. (2), we get the joint entropy
$S_{j}(t)$ for the pure state $\psi(x,t)$ or equivalently it can
be written in the following form ~\cite{Dunkel}
\begin{eqnarray}
    S_{j}(t)&=&-\int dx |\psi(x,t)|^{2}\ln|\psi(x,t)|^{2}-
    \int dp |\tilde{\psi}(p,t)|^2
    \ln |\tilde{\psi}(p,t)|^2-\nonumber\\&-&\ln h^{d}.
\end{eqnarray}
 We find time dependent wave function by means of the Feynman path
integral which has form~\cite{Feynman}
\begin{eqnarray}
K(x'',t'';x',t')&=&\int^{x''=x(t'')}_{x'=x(t')}Dx(t)e^{\frac{i}{\hbar}S[x(t)]}
\nonumber\\&=&\int^{x''}_{x'}Dx(t)e^{\frac{i}{\hbar}\int_{t'}^{t''}L[x,\dot{x},t]dt}.
\end{eqnarray}
The Feynman kernel can be related to the time dependent
Schr\"{o}dinger's wave function
\begin{eqnarray}
K(x'',t'';x',t')=\sum_{n=0}^{\infty}\psi_{n}^{*}(x',t')\psi_{n}(x'',t'').
\end{eqnarray}
The propagator in semiclassical approximation reads
\begin{eqnarray}
K(x'',t'';x',t')=\Big[\frac{i}{2\pi\hbar}\frac{\partial^2}{\partial
x'\partial
x''}S_{cl}(x'',t'';x',t')\Big]^{1/2}e^{\frac{i}{\hbar}S_{cl}(x'',t'';x',t')}.
\end{eqnarray}
The prefactor is often referred to as the Van Vleck-Pauli-Morette
determinant ~\cite{Khandekar,Kleinert}. The $F(x'',t'';x',t')$ is
given by
\begin{eqnarray}
F(x'',t'';x',t')=\Big[\frac{i}{2\pi\hbar}\frac{\partial^2}{\partial
x'\partial x''}S_{cl}(x'',t'';x',t')\Big]^{1/2}.
\end{eqnarray}
\section{Harmonic Oscillator with Time Dependent Mass and
Frequency} The Lagrangian of the harmonic oscillator with
time-dependent frequency and mass are given by
\begin{equation}
    L=\frac{1}{2}m(t)\dot{x}^2-\frac{1}{2}m(t)\omega^{2}(t)x^2,
\end{equation}
where $\omega(t)$ and $m(t)$ are, respectively, the frequency and
mass associated with oscillator, and which are arbitrary real
functions of time. The classical equation of motion for the
Lagrangian is
\begin{equation}
    \ddot{x}+2\frac{\dot{\eta}}{\eta}\dot{x}+\omega^2(t)x=0,
\end{equation}
where $\eta(t)=\sqrt{m}$. The solution to the equation of motion
is given by
\begin{equation}
    x(t)=\frac{\alpha(t)}{\eta(t)}\Big[A \cos\gamma(t)+B
    \sin\gamma(t)\Big],
\end{equation}
where $\frac{\alpha(t)}{\eta(t)}=\rho(t)$, $\gamma(t)$ refers to
the amplitude and phase of classical oscillators and A and B are
constants. The function $\rho(t)$ have to satisfy the following
equation:
\begin{equation}
    \ddot{\rho}+\frac{\dot{m}}{m}\dot{\rho}+\omega^2(t)\rho(t)=\frac{1}{m^2
    \rho^3}.
\end{equation}
The constant A and B in Eq.(13) can be determined by using the
boundary conditions of $x(t')=x'$ and $x(t'')=x''$. The A and B
coefficients yield as
\begin{equation}
    A=\frac{1}{\sin(\gamma''-\gamma')}\Big(\frac{x'}{\rho'}\sin\gamma''-\frac{x''}{\rho''}\sin\gamma'\Big),
\end{equation}
and
\begin{equation}
    B=\frac{1}{\sin(\gamma''-\gamma')}\Big(\frac{x''}{\rho''}\cos\gamma'-\frac{x'}{\rho'}\cos\gamma''\Big).
\end{equation}
Substituting A and B into Eq.(13), the classical path that connects
the point of $(x',t')$ and $(x'',t'')$ can be written as
\begin{equation}
    x_{cl}=\frac{\rho}{\sin(\gamma''-\gamma')}\Big(\frac{x'}{\rho'}\sin(\gamma''-\gamma)-\frac{x''}{\rho''}
    \sin(\gamma''-\gamma)\Big).
\end{equation}
Substituting the classical paths into action function, the
classical action becomes
\begin{eqnarray}
    S_{cl}&=&\frac{m''x''^2}{2}\Big(\frac{\dot{\rho}''}{\rho''}\Big)-
    \frac{m'x'^2}{2}\Big(\frac{\dot{\rho}'}{\rho'}\Big)+\frac{1}{2}\Big(\frac{x''^2}{\rho''^2}+
   \frac{x''^2}{\rho''^2}\Big)\cot(\gamma''-\gamma')-\nonumber\\&-&\frac{x''x'}{\rho''\rho'}\csc(\gamma''-\gamma').
\end{eqnarray}
By substituting the above action into Eq.(10), the factor can be
obtained as
\begin{eqnarray}
F(x'',t'';x',t')=\Big[\frac{1}{2i\pi\hbar\rho'\rho''\sin(\gamma''-\gamma')}\Big]^{1/2}.
\end{eqnarray}
Therefore the propagator for the harmonic oscillator with a time
dependent mass and frequency can be obtained by
\begin{eqnarray}
K(x'',t'';x',t')&=&\Big[\frac{1}{2i\pi\hbar\rho'\rho''\sin(\gamma''-\gamma')}\Big]^{1/2}
\exp\Big[\frac{i}{2\hbar}\Big[\frac{m''x''^2}{2}\Big(\frac{\dot{\rho}''}{\rho''}\Big)-\nonumber\\&-&
    \frac{m'x'^2}{2}\Big(\frac{\dot{\rho}'}{\rho'}\Big)\Big]\Big]
    \exp\Big[\frac{i}{2\hbar\sin(\gamma''-\gamma')}\Big(\frac{x''^2}{\rho''^2}
    +\frac{x'^2}{\rho'^2}\Big)\times\nonumber\\&\times&\cos(\gamma''-\gamma')-\frac{2x''x'}{\rho'\rho''}\Big].
\end{eqnarray}
By the use of the Mehler-formula
\begin{eqnarray}
    e^{-(x^2+y^2)/2}\sum_{n=0}^{\infty}\frac{1}{n!}(\frac{z}{2})^2H_{n}(x)H_{n}(y)&=&\frac{1}{\sqrt{1-z^2}}
    \exp[\frac{4xyz}{2(1-z^2)}-\nonumber\\&-&\frac{(x^2+y^2)(1+z^2)}{2(1-z^2)}],
\end{eqnarray}
where $H_{n}$ is Hermite polynomials, we can write the Feynman
kernel in form
\begin{eqnarray}
K(x'',t'';x',t')&=&\frac{1}{2^n
n!}\Big[\frac{1}{2i\pi\hbar\rho'\rho''\sin(\gamma''-\gamma')}\Big]^{1/2}
\exp\Big[\frac{i}{2\hbar}\Big[\frac{m''x''^2}{2}\Big(\frac{\dot{\rho}''}{\rho''}\Big)-\nonumber\\&-&
    \frac{m'x'^2}{2}\Big(\frac{\dot{\rho}'}{\rho'}\Big)\Big]\Big]
    \exp\Big[-\frac{1}{2\hbar}\Big(\frac{x''^2}{\rho''^2}
    +\frac{x'^2}{\rho'^2}\Big)\Big]\sum H_{n}\Big[\sqrt{\frac{1}{\hbar}}\frac{x''}{\rho''}\Big]
    \times\nonumber\\&\times&H_{n}\Big[\sqrt{\frac{1}{\hbar}}\frac{x'}{\rho'}\Big]\exp\Big(-i\gamma(n+\frac{1}{2})\Big).
\end{eqnarray}
Comparing the kernel in Eq. (22) with Eq. (8), the wave function for
harmonic oscillator with time dependent mass and frequency can be
found as
\begin{eqnarray}
 \Psi_{n}(x,t)&=&\exp[i\alpha_{n}(t)]\Big[\frac{1}{\pi^{1/2}\hbar^{1/2}n!2^{n}\rho}\Big]^{1/2}
 \exp\Big[\frac{i m(t)}{2\hbar}\Big(\frac{\dot{\rho}}{\rho}+\frac{i}{m(t)\rho^{2}}\Big)x^{2}\Big]
 \times\nonumber\\&\times&H_{n}\Big[\Big(\frac{1}{\hbar}\Big)^{1/2}\frac{x}{\rho}\Big],
\end{eqnarray}
where the phase functions $\alpha_{n}(t)$ are described by
\begin{eqnarray}
\alpha_{n}(t)=-\Big(n+\frac{1}{2}\Big)\int_{0}^{t}\frac{1}{m(t')\rho^{2}}dt'.
\end{eqnarray}
This result is also briefly obtained before by using the Feynman
Path Integral method in Ref.\cite{Khandekar}.  It is also in
agreement with ones obtained in Refs. ~\cite{Pedrosa} and
~\cite{Ciftja}. The time dependent wave function of ground state
is
\begin{eqnarray}
 \Psi_{0}(x,t)&=&\exp[i\alpha_{0}(t)]\Big[\frac{1}{\pi^{1/2}\hbar^{1/2}\rho}\Big]^{1/2}
 \exp\Big[\frac{i
 m(t)}{2\hbar}\Big(\frac{\dot{\rho}}{\rho}+\frac{i}{m(t)\rho^{2}}\Big)x^{2}\Big].
\end{eqnarray}
Note that when $m(t)\rightarrow m$, $\omega(t)\rightarrow\omega_0$
and $\rho(t)\rightarrow\rho_0=const=1/\sqrt{m\omega_0}$, the
solution in Eq.(23) becomes the solution for the time independent
harmonic oscillator of mass m and frequency $\omega_0$.
\subsection{Time dependent mass and constant frequency}
\begin{figure}[htbp]
\centering \epsfig{file=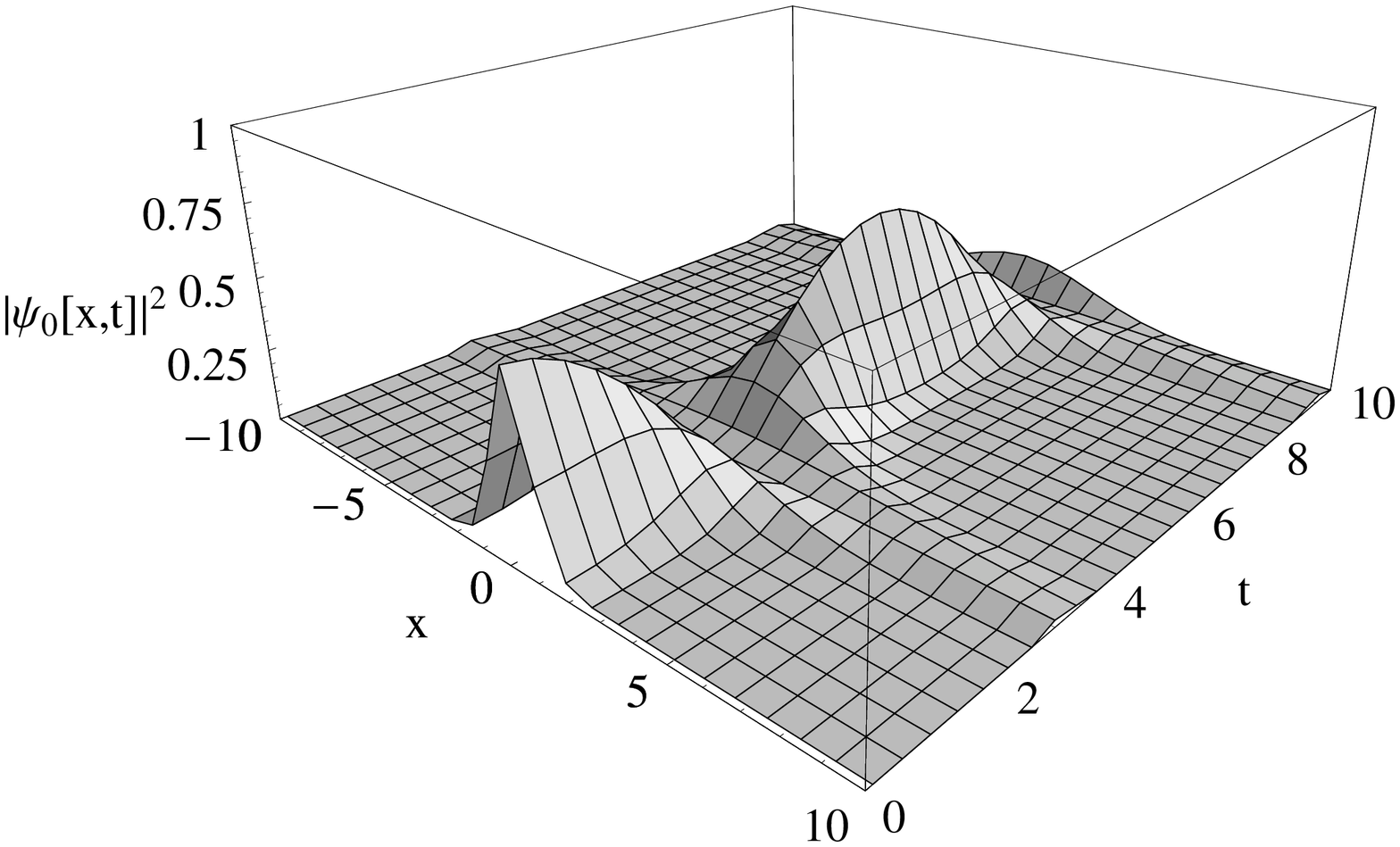, width=12cm,height=12cm}
\caption{$|\Psi_{0}(x,t)|^{2}$ versus time and coordinate for
strongly pulsating mass and constant frequency.}\label{eps1}
\end{figure}

\begin{figure}[htbp]
\centering \epsfig{file=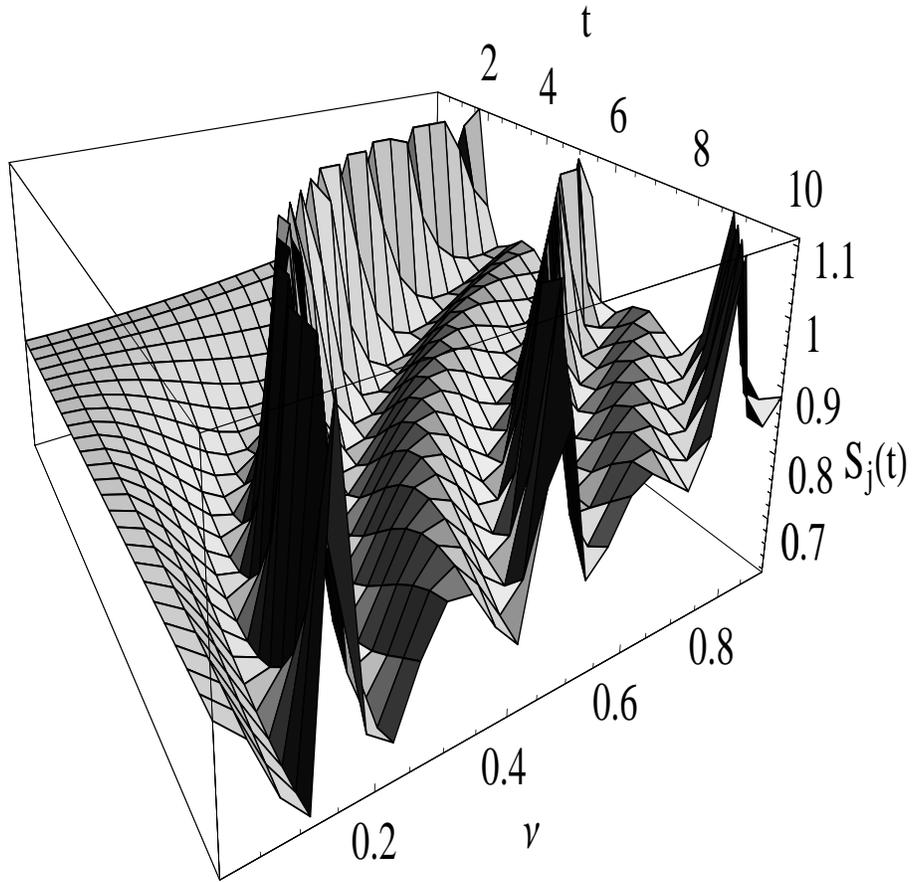, width=12cm,height=12cm}
\caption{The 3D graph of joint entropy of harmonic oscillator with
time dependent mass and constant frequency.}\label{eps2}
\end{figure}

\begin{figure}[htbp]
\centering \epsfig{file=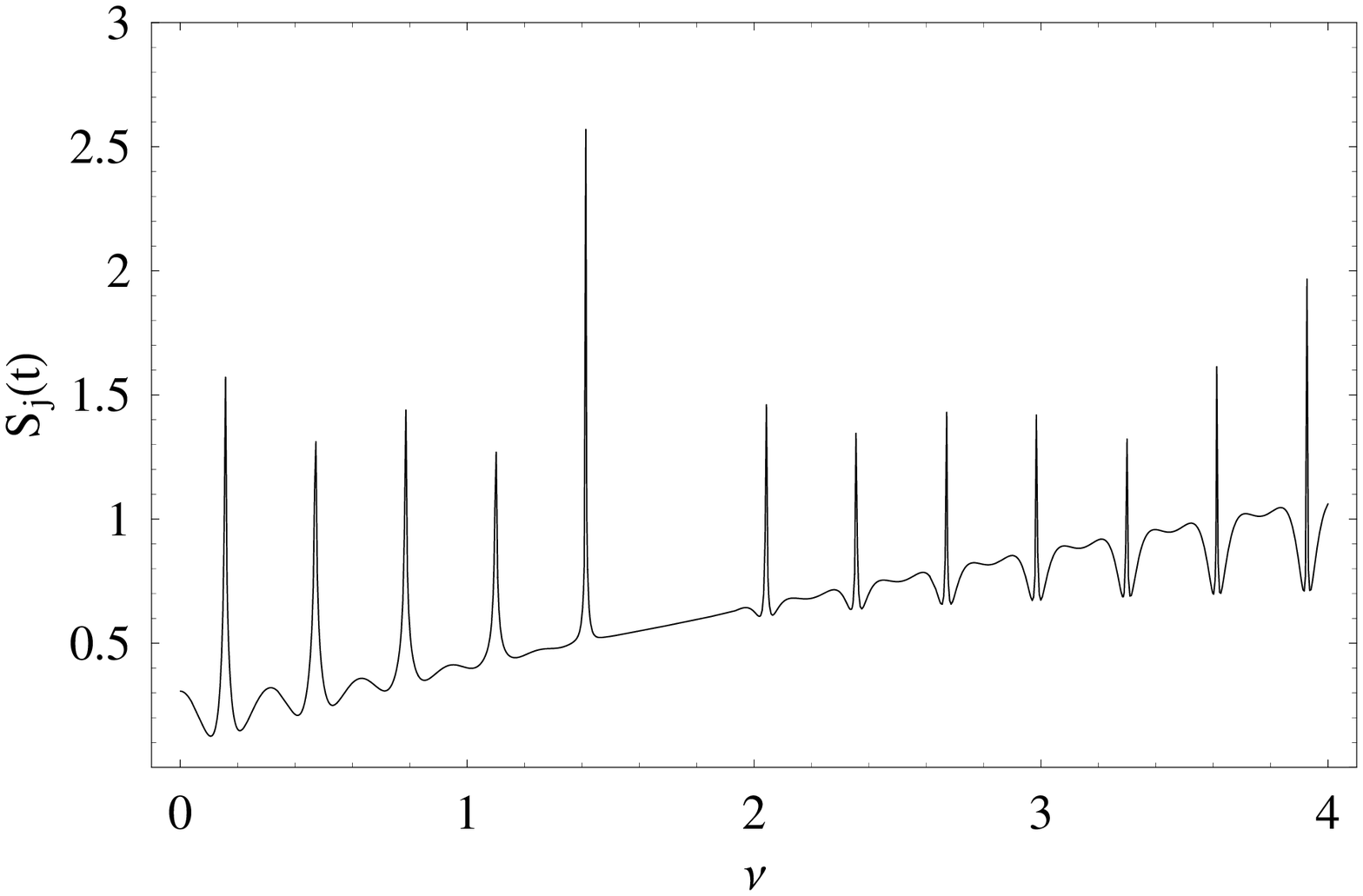, width=12cm,height=12cm}
\caption{The joint entropy of harmonic oscillator with time
dependent mass and constant frequency versus small  $\nu$.
}\label{eps3}
\end{figure}

\begin{figure}[htbp]
\centering \epsfig{file=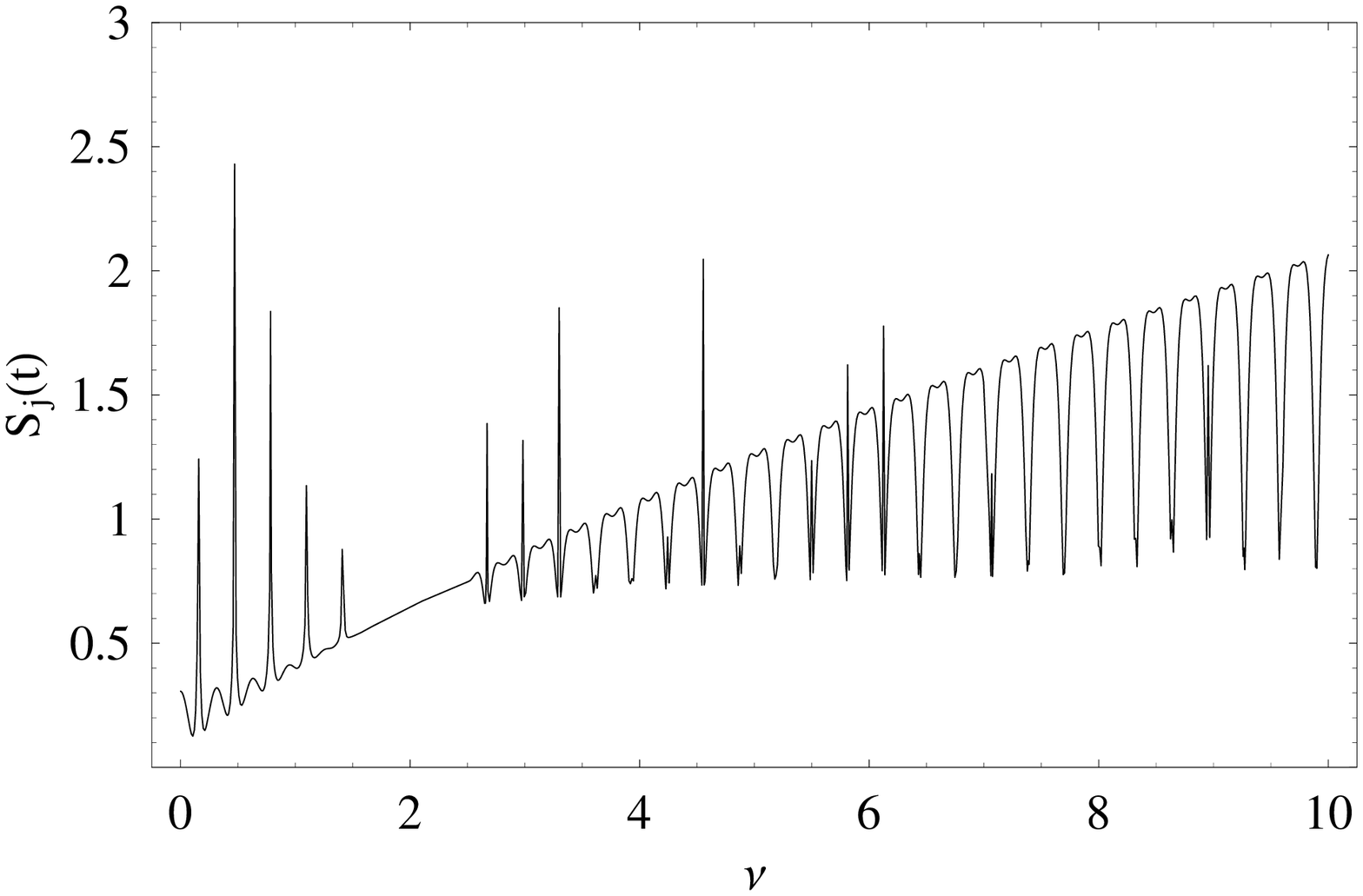, width=12cm,height=12cm}
\caption{The joint entropy of harmonic oscillator with time
dependent mass and constant frequency versus large
$\nu$.}\label{eps4}
\end{figure}

In this calculation, we choose harmonic oscillator constant
frequency and time dependent mass which is called pulsating mass.
For instance, it has been shown that the Lagrangian describing the
problem of a Fabry-Perot cavity in contact with a heat reservoir
assumes the form of constant frequency and time dependent
mass~\cite{Colegrave}. The time dependence of the mass  can be taken
as $m(t)=m \cos^2(\nu t)$ where $\nu$ is described as frequency of
pulsating mass. The Lagrangian of the system is
\begin{equation}
    L=\frac{1}{2}m \cos^2(\nu t)\dot{x}^2-\frac{1}{2}m \cos^2(\nu
    t)\omega^{2}x^2.
\end{equation}
The kernel for a harmonic oscillator with strongly pulsating mass
can be obtained from Eq. (20) as
\begin{eqnarray}
K(x'',t'';x',t')&=&\Big[\frac{m\Omega\cos\nu t'\cos\nu
t''}{2i\pi\hbar\sin\Omega(t''-t')}\Big]^{1/2}
\exp\Big[\frac{im\nu}{2\hbar}\Big(\cos^2\nu\tan\nu
t''x''^2-\nonumber\\&-&\cos^2\nu t'\tan\nu t'x'^2\Big)\Big]
\exp\Big[\frac{im\Omega}{2\hbar\sin(\Omega t''-t')}\Big(\cos^2\nu
t''x''^2+\nonumber\\&+&\cos^2\nu
t'x'^2\Big)\cos\Omega(t''-t')-2\cos\nu t'\cos\nu t''x''x'\Big],
\end{eqnarray}
where we use $\Omega^2=\omega^2+\nu^2$.
 Using Eq. (21) Mehler formula, and then the time dependent wave function was obtained as
\begin{eqnarray}
\Psi_{n}(x,t)&=&\exp[-i\beta_{n}(t)]\Big[\frac{m(t)\Omega}{\pi\hbar
n!^2 2^{2n}}\Big]^{1/4}
 \exp\Big[\frac{im(t)}{2\hbar}\Big(\nu\tan\nu t+i\Omega\Big)x^{2}\Big]
 \times\nonumber\\&\times&H_{n}\Big[\Big(\frac{m(t)\Omega}{\hbar}\Big)^{1/2}x\Big],
\end{eqnarray}
where  the phase functions $\beta_{n}(t)$ is described as
\begin{eqnarray}
\beta_{n}(t)=(n+1/2)\Omega t.
 \end{eqnarray}
Note that when $\nu\rightarrow 0$ and $m(t)\rightarrow m=const$, the
solution in Eq.(28) reduces to the solution for the time independent
harmonic oscillator of mass m and frequency $\omega_0$.  The ground
state wave function yields as
\begin{eqnarray}
\Psi_{0}(x,t)&=&\exp[-i(1/2)\Omega
t]\Big[\frac{m(t)\Omega}{\pi\hbar}\Big]^{1/4}
 \exp\Big[\frac{im(t)}{2\hbar}\Big(\nu\tan\nu
 t+i\Omega\Big)x^{2}\Big].
\end{eqnarray}
The density of probability in coordinate space is
\begin{eqnarray}
|\Psi_{0}(x,t)|^2&=&\Big[\frac{m(t)\Omega}{\pi\hbar}\Big]^{1/2}
 \exp\Big[-\frac{m(t)\Omega}{\hbar}x^{2}\Big].
\end{eqnarray}
In Fig.\ref{eps1}, the density of probability in coordinate space is
shown. The density of probability in momentum space can be easily
calculated as
\begin{eqnarray}
|\Psi_{0}(p,t)|^2&=&\Big[\frac{\Omega}{\hbar\pi m(\nu^2\tan^2\nu
t+\Omega^2)}\Big]^{1/2}
 \exp\Big[-\frac{p^{2}\Omega}{m\hbar(\nu^2\tan^2\nu
 t+\Omega^2)}\Big].
\end{eqnarray}
 The joint entropy for ground state from Eq. (6) becomes

\begin{eqnarray}
S_{j}(t)&=&\frac{1}{2}\Big[\ln\Big(\frac{e^2}{4}\Big)+\ln\Big(\frac{1}{\pi\hbar\sqrt{m(t)\Omega}}\Big)
\nonumber\\&+&\sqrt{m(t)\Omega}\Big(1-\ln\frac{\Omega}{\pi\hbar(\nu^2\tan^2\nu
t+\Omega^2)}\Big)\Big].
\end{eqnarray}
The joint entropy of harmonic oscillator with time dependent mass
and constant frequency is plotted in Fig.\ref{eps2}. The small $\nu$
and large $\nu$ cases are shown in Fig.\ref{eps3} and
Fig.\ref{eps4}, respectively.
\subsection{Time dependent frequency and constant mass}
\begin{figure}[htbp]
\centering \epsfig{file=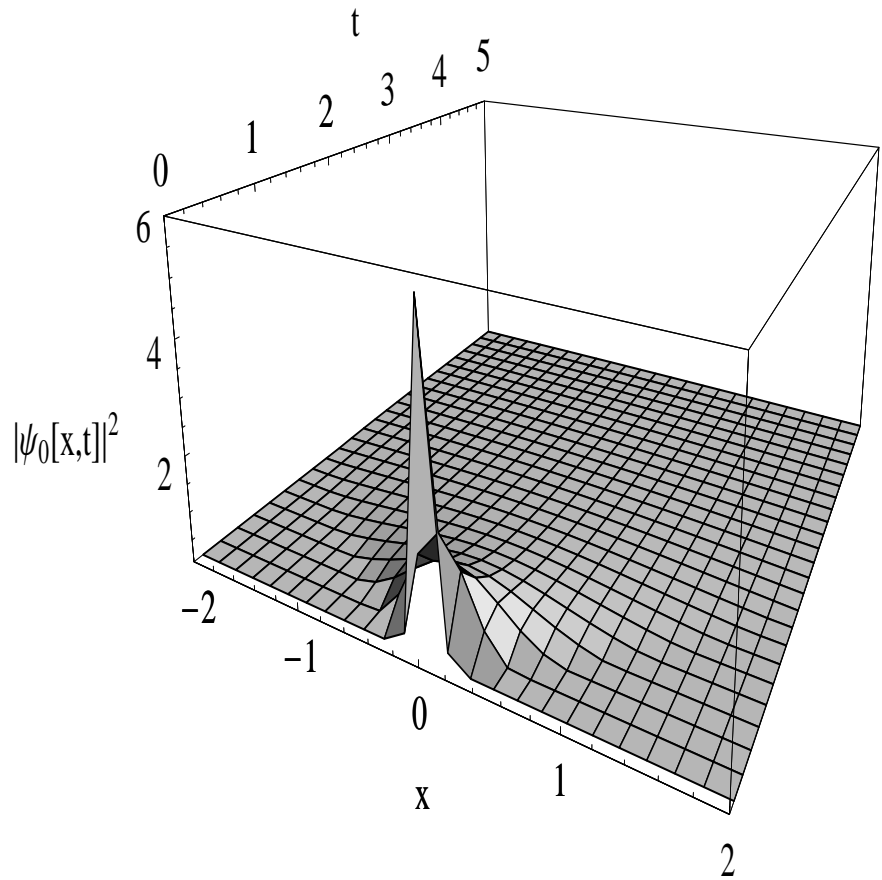, width=12cm,height=12cm}
\caption{$|\Psi_{0}(x,t)|^{2}$ versus  time and coordinate at
constant mass and inverse square time frequency.}\label{eps5}
\end{figure}

\begin{figure}[htbp]
\centering \epsfig{file=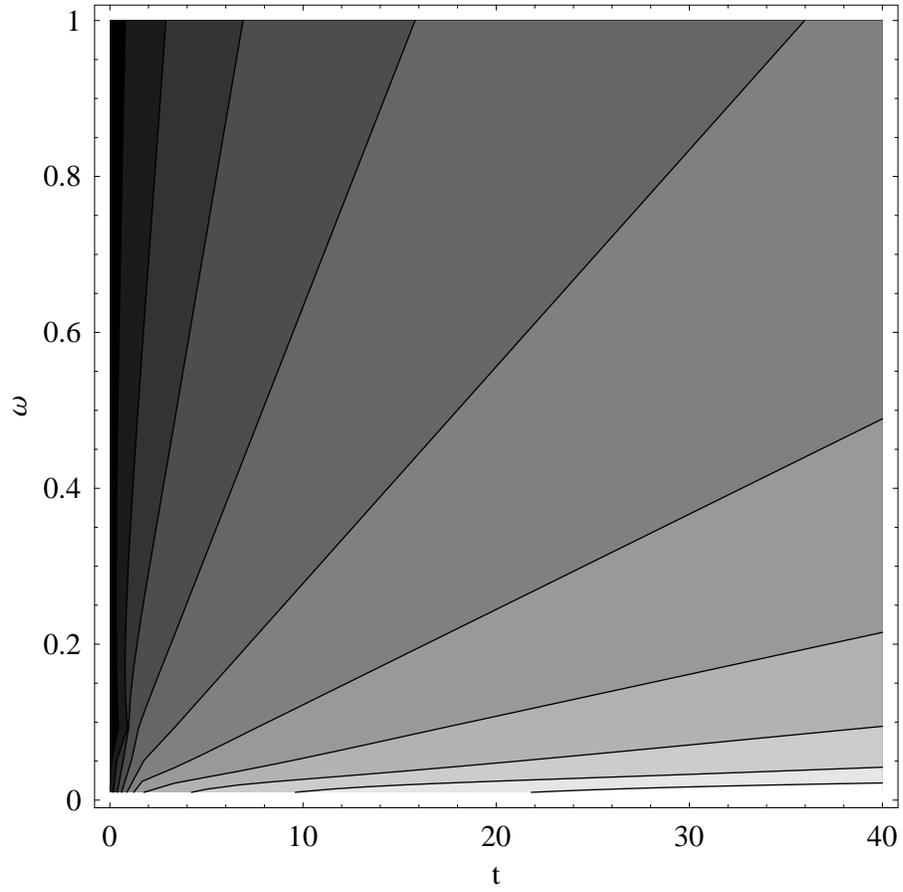, width=12cm,height=12cm}
\caption{The counter graph of the joint entropy of Harmonic
oscillator with  inverse square time dependent frequency and
constant mass. }\label{eps6}
\end{figure}

\begin{figure}[htbp]
\centering \epsfig{file=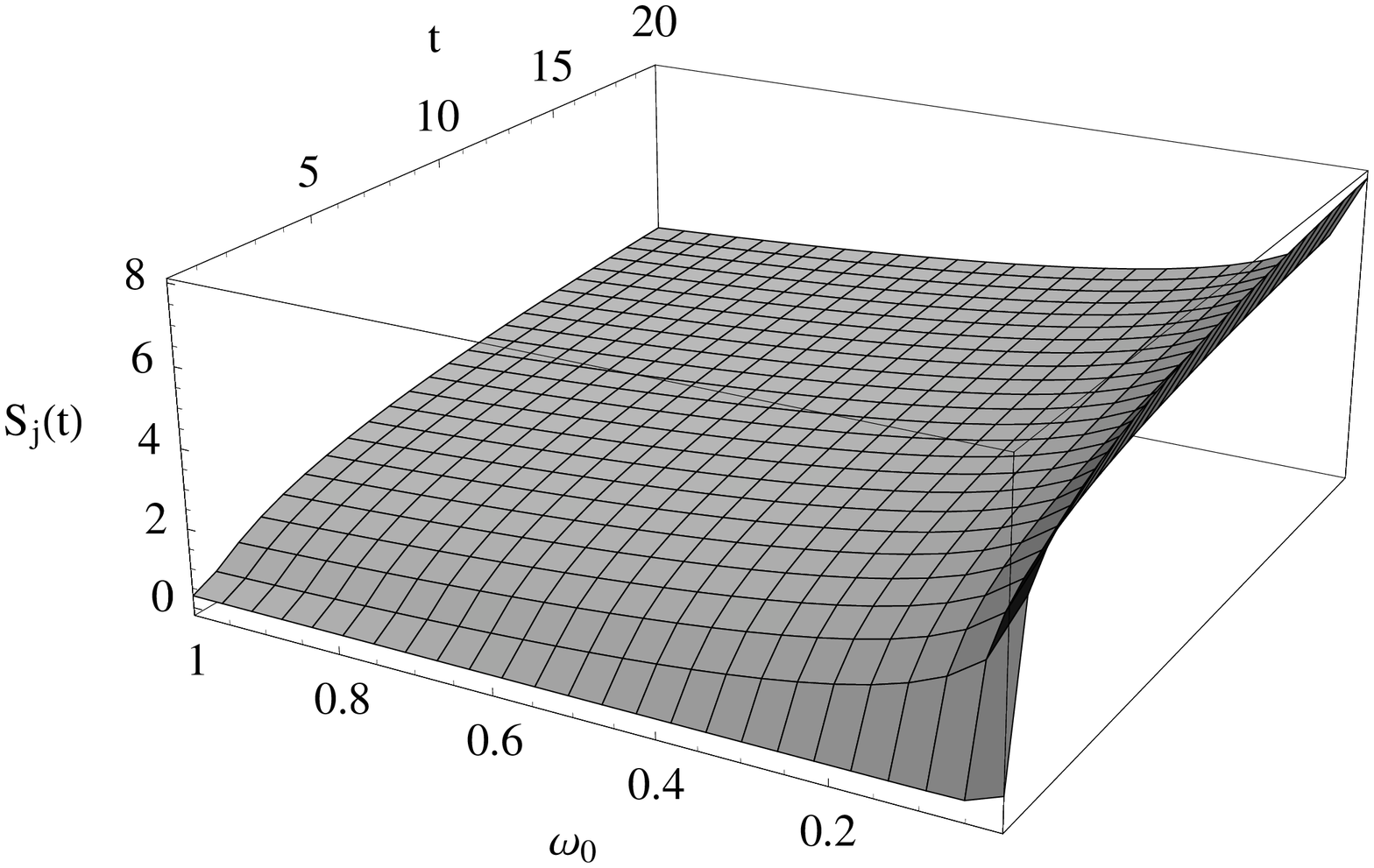, width=12cm,height=12cm}
\caption{The 3D graph of the joint entropy of harmonic oscillator
with inverse square time dependent frequency}\label{eps7}
\end{figure}

The Lagrangian of the harmonic oscillator with time-dependent
frequency is given by
\begin{equation}
    L=\frac{1}{2}m \dot{x}^2-\frac{1}{2}m\omega(t)^{2}x^2.
\end{equation}
 The time dependent frequency is described by
\begin{equation}
\omega(t)=\frac{\omega_0}{t^2}.
\end{equation}
In this case, we defined the following quantities
\begin{equation}
\rho(t)=\frac{1}{\sqrt{m\omega(t)}},
\end{equation}
and
\begin{equation}
\gamma(t)=-\frac{\omega_0}{t}.
\end{equation}
By substituting Eqs. (35), (36) and (37) into the general kernel Eq.
(20), the kernel for a harmonic oscillator with the inverse square
time dependent frequency can be derived as
\begin{eqnarray}
K(x'',t'';x',t')&=&\Big[\frac{m\omega_0}{2i\pi\hbar t'
t''\sin\omega_0\Big(\frac{t''-t'}{t't''}\Big)}\Big]^{1/2}\exp\Big[\frac{i}{2\hbar}\Big(\frac{mx''^2}{t''}-
\frac{mx'^2}{t'}\Big)\Big]\times\nonumber\\&\times&\exp\Big[\frac{im\omega_0}{2\hbar
\sin\omega_0\Big(\frac{t''-t'}{t't''}\Big)}\Big[\Big(\frac{x''^2}{t''^2}+\frac{x'^2}{t'^2}\Big)
\cos\omega_0\Big(\frac{t''-t'}{t't''}\Big)-\frac{2x'x''}{t't''}\Big].
\end{eqnarray}
Using Eq. (21) Mehler formula, we obtain the kernel
\begin{eqnarray}
K(x'',t'';x',t')&=&\Big[\frac{m\omega_0}{\pi\hbar t'
t''}\Big]^{1/2}\exp\Big[\frac{i}{2\hbar}\Big(\frac{mx''^2}{t''}-
\frac{mx'^2}{t'}\Big)\Big]\times\nonumber\\&\times&\exp\Big[-\frac{m\omega_0}{2\hbar
}\Big[\Big(\frac{x''^2}{t''^2}+\frac{x'^2}{t'^2}\Big)\Big]\sum_{0}^{\infty}H_n\Big(\sqrt{\frac{m\omega_0}{\hbar}}
\frac{x''}{t''}\Big)H_n\Big(\sqrt{\frac{m\omega_0}{\hbar}}
\frac{x'}{t'}\Big)\times\nonumber\\&\times&\frac{\exp(\frac{i\omega_0}{t}(n+1/2))}{2^n
n!}.
\end{eqnarray}
Using the Eq. (8), the time dependent wave function was derived as
\begin{eqnarray}
\Psi_n(x,t)&=&\frac{1}{\sqrt{2^nn!t}}\Big[\frac{m\omega_0}{\pi\hbar}\Big]^{1/4}
\exp\Big(\Big(n+\frac{1}{2}\Big)\frac{i\omega_0}{t}\Big)\exp\Big[\frac{im}{2\hbar
t}\Big(1+\frac{i\omega_0}{t}\Big)x^2\Big]\times\nonumber\\&\times&
H_n\Big(\sqrt{\frac{m\omega_0}{\hbar}} \frac{x}{t}\Big).
\end{eqnarray}
The time dependent wave function for ground state is given by
\begin{eqnarray}
\Psi_0(x,t)&=&\Big[\frac{m\omega_0}{\pi\hbar t^2}\Big]^{1/4}
\exp\Big(\frac{i\omega_0}{2t}\Big)\exp\Big[\frac{im}{2\hbar
t}\Big(1+\frac{i\omega_0}{t}\Big)x^2\Big].
\end{eqnarray}
The density of probability in coordinate space is
\begin{eqnarray}
|\Psi_0(x,t)|^2&=&\Big[\frac{m\omega_0}{\pi\hbar t^2}\Big]^{1/2}
\exp\Big[-\frac{m\omega_0}{\hbar t^2}x^2\Big].
\end{eqnarray}
The density of probability for coordinate space is shown in
Fig.\ref{eps5}. The density of probability in momentum space can be
easily calculated as
\begin{eqnarray}
|\Psi_0(p,t)|^2&=&\Big[\frac{\omega_{0}t^2}{(\omega_{0}^2+t^2)\hbar\pi
m}\Big]^{1/2}\exp\Big[-\frac{p^2\omega_{0}t^{2}}{m\hbar(\omega_{0}^2+t^2)}\Big].
\end{eqnarray}
Note that when $t\rightarrow 1$ and
$\omega(t)\rightarrow\omega_0=const$, the solution of Eqs.(42) and
(43) become the solution for the time independent harmonic
oscillator of mass m and frequency $\omega_0$.
 The joint entropy for ground state from Eq. (6) becomes
\begin{eqnarray}
S_{j}(t)=\ln\Big[\Big(\frac{e}{2}\Big)\Big(\frac{\omega_{0}^2+t^2}{\omega_{0}^2}\Big)^{1/2}\Big].
\end{eqnarray}
The contour and joint entropy of harmonic oscillator with inverse
square time dependent frequency are shown in Figs. \ref{eps6} and
\ref{eps7}, respectively. It is important that Eq. (44) is in
agreement with following general inequality for the joint entropy:
\begin{equation}
    S_{j}(t)\geq\ln(\frac{e}{2}).
\end{equation}
originally derived by Leipnik for arbitrary one-dimensional
one-particle wave functions\cite{Leipnik,Dunkel}.
\section{Conclusion}
We have investigated the joint entropy for explicit time dependent
solution of one-dimensional harmonic oscillators with time dependent
frequency and mass. The time dependent wave function is obtained by
means of Feynman Path integral technique. In the time dependent
strongly pulsating mass, harmonic oscillator with constant frequency
case, we have found that the joint entropy fluctuated with time and
frequency. It is seen that, the joint entropy has harmonic behavior.
This result indicates that the information periodically transfers
between harmonic oscillators with strongly pulsating masses. On the
other hand, inverse square time frequency case, the joint entropy of
harmonic oscillator with time dependent frequency shows remarkable
monotonically increase with time. In the
$\omega(t)\rightarrow\omega=constant$ and $\nu\rightarrow 0$, case
theses results agree with time independent harmonic oscillator. It
also depends on choice of initial frequency as seen in
Fig.~\ref{eps7}.

\section{Acknowledgements}

This research was partially supported by the Scientific and
Technological Research Council of Turkey.

\newpage
\newpage
\newpage
\newpage
\newpage
\newpage
\newpage
\


\begin{thebibliography}{20}

\bibitem{ekrem}
E. Aydiner, C. Orta and R.Sever, E-print:quant-ph/0602203
\bibitem{Zurek}
W.H. Zurek, Phys. Today \textbf{44}(10), 36 (1991).
\bibitem{Omnes}
R. Omnes, Rev. Mod. Phys. \textbf{64}, 339 (1992).
\bibitem{Anastopoulos}
C. Anastopoulos, Ann. Phys. \textbf{303}, 275 (2003).
\bibitem{Leipnik}
R. Leipnik, Inf. Control. \textbf{2}, 64 (1959).
\bibitem{Dodonov}
V.V.Dodonov, J.Opt. B: Quantum Semiclassical Opt. \textbf{4}, S98
(2002).
\bibitem{Trigger}
S. A. Trigger, Bull. Lebedev Phys. Inst. \textbf{9}, 44 (2004).
\bibitem{Dunkel}
J. Dunkel and S. A. Trigger, Phys. Rev.A\textbf{71}, 052102 (2005).
\bibitem{Garbaczewski}
P. Garbaczewski, Phys. Rev. A \textbf{72}, 056101 (2005).
\bibitem{Dandas}
C. M. A. Dantas, I. A . Pedrosa and B. Baseia, Phys. Rev.A
\textbf{45}, 3 (1992).
\bibitem{Abdalla}
R. K. Colegrave and M. S. Abdalla, Opt. Acta \textbf{28}, 495
(1981).
\bibitem{Lemos}
N. A. Lemos and C. P. Natividade, Nuovo Cimento \textbf{399}, 211
(1989).
\bibitem{Ben}
Y. Ben-Aryeh and A. Mann, Pyhs. Rev. A \textbf{32}, 552 (1985).
\bibitem{Born}
M. Born, Z. Phys. \textbf{40}, 167 (1926).
\bibitem{Feynman}
R.P. Feynman, A. R. Hibbs, Quantum Mechanics and Path Integrals,
McGraw-Hill, USA (1965).
\bibitem{Khandekar}
D.C. Khandekar, S.V. Lawande, K.V. Bhagwat, Path-Integral Methods
and Their Applications, World Scientific, Singapore (1993).
\bibitem{Kleinert}
H. Kleinert, Path Integrals in Quantum Mechanics, Statistics,
Polymer Physics, and Financial Markets, World Scientific, 3rd
Edition (2004).
\bibitem{Pedrosa}
I.A. Pedrosa, Phys. Rev. A \textbf{55}, 3219 (1997).
\bibitem{Ciftja}
O. Ciftja, J. Phys A \textbf{32},6385 (1999).
\bibitem{Colegrave}
R. K. Colegrave and M. S. Abdalla, J. Phys. A \textbf{15},
1549(1950).
\end{thebibliography}
\end{document}